\documentclass[10pt]{article}

\usepackage{amsmath}
\usepackage{amssymb}

\usepackage{graphicx}

\usepackage{cite}

\usepackage{color} 

\usepackage[labelfont=bf,labelsep=period,justification=raggedright]{caption}

\makeatletter
\renewcommand{\@biblabel}[1]{\quad#1.}
\makeatother

\date{}

\begin{document}

% Title must be 150 words or less
\begin{flushleft}
{\Large
\textbf{Active dendrites enhance neuronal dynamic range}
}
% Insert Author names, affiliations and corresponding author email.
\\
Leonardo L. Gollo$^{1,2,\ast}$, 
Osame Kinouchi$^{3}$,
Mauro Copelli$^{1}$
\\
\bf{1} Laborat\'orio de F{\'\i}sica 
Te\'orica e Computacional, Departamento
de F{\'\i}sica, Universidade Federal de Pernambuco, Recife,
PE, Brazil
\\
\bf{2} 
IFISC, Instituto de F\'{\i}sica Interdisicplinar y Sistemas Complejos (CSIC-UIB), Campus Universitat Illes Balears, E-07122 Palma de Mallorca, Spain
\\
\bf{3} Faculdade de Filosofia, Ci\^encias e Letras de
Ribeir\~ao Preto, Universidade de S\~ao Paulo,
Ribeir\~ao Preto, SP, Brazil
\\
$\ast$ E-mail: leonardo@ifisc.uib-csic.es
\end{flushleft}

% Please keep the abstract between 250 and 300 words
\section*{Abstract}

Since the first experimental evidences of active conductances in
dendrites, most neurons have been shown to exhibit dendritic
excitability through the expression of a variety of voltage-gated ion
channels. However, despite experimental and theoretical efforts
undertaken in the last decades, the role of this excitability for some
kind of dendritic computation has remained elusive. Here we show that,
owing to very general properties of excitable media, the average
output of a model of an active dendritic
tree is a highly non-linear function of its
afferent rate, attaining extremely large dynamic ranges (above 50
dB). Moreover, the model yields double-sigmoid response functions as
experimentally observed in retinal ganglion cells. We claim that
enhancement of dynamic range is the primary functional role of active
dendritic conductances. We predict that neurons with larger dendritic
trees should have larger dynamic range and that blocking of active
conductances should lead to a decrease of dynamic range.

% Please keep the Author Summary between 150 and 200 words
% Use first person. PLoS ONE authors please skip this step. 
% Author Summary not valid for PLoS ONE submissions.   
\section*{Author Summary}

Most neurons present cellular tree-like extensions known as dendrites,
which receive input signals from synapses with other cells.  Some
neurons have very large and impressive dendritic arbors. What is the
function of such elaborate and costly structures?  The functional role
of dendrites is not obvious because, if dendrites were an electrical
passive medium, then signals from their periphery could not influence
the neuron output activity. Dendrites, however, are not passive, but
rather active media that amplify and support pulses (dendritic
spikes). These voltage pulses do not simply add but can also
annihilate each other when they collide. To understand the net effect
of the complex interactions among dendritic spikes under massive
synaptic input, here we examine a computational model of excitable
dendritic trees. We show that, in contrast to passive trees, they have
a very large dynamic range, which implies a greater capacity of the
neuron to distinguish among the widely different intensities of input
which it receives. Our results provide an explanation to the
concentration invariance property observed in olfactory processing,
due to the very similar response to different inputs. In addition, our
modeling approach also suggests a microscopic neural basis for the
century old psychophysical laws.

\section*{Introduction}

One of the distinctive features of many neurons is the presence of
extensive dendritic trees. Much experimental and computational work
has been devoted to the description of morphologic and dynamic aspects
of these neural processes~\cite{Stuart99}, in special after the
discovery of dendritic active
conductances~\cite{Eccles58,Gulledge04,London05}. Several proposals
have been made about possible computational functions associated to
active dendrites, such as the implementation of biological logic gates
and coincidence detectors~\cite{Koch,Squire}, learning signaling via
dendritic spikes~\cite{GoldingStaffSpruston02} or an increase in the
learning capacity of the neuron~\cite{PoiraziMel2001}. However, it is
not clear whether such mechanisms are robust in face of the noisy and
spatially distributed character of incoming synaptic input, as well as
the large variability in morphology and dendritic sizes.

Here we propose to view the dendritic tree not as a computational
device, an exquisitely designed ``neural microchip''~\cite{Squire}
whose function could be dependent on an improbable fine tuning of
biological parameters (such as delay constants, arborization size,
etc), but rather as a spatially extended excitable 
system~\cite{Lindner04} whose robust collective properties may have
been progressively exapted to perform other biological functions. Our
intention is to provide a simpler hypothesis about the functional role
of active dendrites, which could be experimentally tested against
other proposals.

We study a model where the excitable dynamics is simple, but the
dendritic topology is faithfully reproduced by means of a binary tree
with a large number of excitable branchlets. Most importantly,
branchlets are activated stochastically (at some rate), so that the
effects of the nonlinear interactions among dendritic spikes can be
assessed. We study how the geometry of such a spatially extended
excitable system boosts its ability to perform non-linear signal
processing on incoming stimuli. We show that excitable trees naturally
exhibit large dynamic ranges --- above 50~dB. In other words, the
neuron could handle five orders of magnitude of stimulus intensity,
even in the absence of adaptive mechanisms.  This performance is one
hundred times better than what was previously observed in other
network topologies~\cite{Kinouchi06a,Wu07}.

Such a high performance seems to be characteristic of branched (tree)
structures. We believe that these findings provide important clues
about the possible functional roles of active dendrites, thus
providing a theoretical background~\cite{London05} on the cooperative
behavior of interacting branchlets. We observe in the model the
occurrence of dendritic spikes similar to those already observed
experimentally and recently related to synaptic
plasticity~\cite{GoldingStaffSpruston02}. Here, however, such spikes
are just an inevitable consequence of the excitable dynamics and we
propose that even dendritic trees without important plasticity
phenomena (like those of some sensory neurons) could benefit from
active dendrites from the point of view of enlargement of its
operational range.
 
Our results also suggest that, under continuous synaptic bombardment,
dendritic spikes could be responsible for another unintended
prediction of the model, namely, that the neuron transfer function
needs not to be simply a Hill-like saturating curve; rather, a
double-sigmoid behavior may appear (as observed experimentally in
retinal ganglion cells~\cite{Deans02}). The model further predicts
that:
\begin{itemize}
\item  the neuron average activity depends mainly on
the rate of branchlet activation, reflecting in a robust way the
afferent input, and not on the total number of
branchlets present in the tree, which is highly dependent on
accidental morphological details; 
\item the size of the dendritic
arbor, or rather the number of bifurcations of the tree, affects in a
specific manner the neuronal dynamic range.
\end{itemize}

So, why do neurons have active dendrites? As a short answer, we
propose that neurons are the only body cells with large dendrites
because they need to work with a large stimulus range. Owing to the
enormous number of afferent synapses and the large variability of
input rates, highly arborized and {\it active} dendrites are crucial
to enhance the dynamic range of a neuron, in a way not accounted for
by passive cable theory and biophysical neuron models with few
compartments (reduced models)~\cite{RothHausser01}. Other phenomena,
such as backpropagating spikes, could have been later exapted to more
complex functional roles.  One should, however, consider first a
generic property of extended excitable media: that, due to the
creation and annihilation of non-linear pulses, the input-output
transfer function of such media is necessarily highly non-linear, with
a very large dynamic range as compared with that of a passive medium.

% You may title this section "Methods" or "Models". 
% "Models" is not a valid title for PLoS ONE authors. However, PLoS ONE
% authors may use "Analysis" 
\section*{Methods}

\subsection*{Modeling active dendrites}

In computational neuroscience, the behavior of an active neuronal
membrane traditionally is modeled by coupled differential equations
which represent the dynamics of its electric potential and gating
variables related to the ionic conductances. This modeling strategy
was then further extended by detailing the dendritic tuft through a
compartmental approach~\cite{Rall64}. Motivated by the abundant
evidence that dendrites have active ion channels that can support
non-linear summation and dendritic spikes~\cite{Stuart99,Squire}, this
line of research currently aims at examining the possibility that
these extensive tree-shaped neuronal regions may be the stage for some
kind of ``dendritic
computation''~\cite{London05,Koch,JohnstonReviewDendrites96}.

Many efforts within this framework of biophysical modeling have
been devoted to unveiling the conditions under which the regenerative
properties of dendritic active channels may be unleashed to generate a
nonlinear excitation (e.g. at the level of a single
spine~\cite{SegevRall1988} or upon temporally synchronized and locally
strong input at the level of a branchlet~\cite{Mel93}). Nonlinear
cable theory can further help predict whether and how a single
dendritic spike will propagate along the branches, for instance
highlighting the relative importance of a given channel type for the
propagation of action potentials~\cite{Migliore99}.  Detailed
biophysical models also correctly predicts e.g. that two
counter-propagating dendritic spikes annihilate each other upon
collision~\cite{RumseyAbbott,RoyerMiller07} (instead of summing), but
this is true for most -- if not all -- extended excitable media.
However, at the present state of the art of neuronal simulations,
biophysical modeling may not necessarily be the approach best suited
for addressing the much more difficult question of what happens when
many dendritic spikes interact, specially in a more natural scenario
where they would be continuously created at different points of the
dendritic tree at some stochastic rate.

Understanding the net effect of the creation and annihilation of
dendritic nonlinear excitations under massive spatio-temporal patterns
of synaptic input requires 1) knowledge of the key properties of these
excitations and their interactions (which cable theory gives us) and
2) a theoretical framework which addresses the resulting collective
behavior. We therefore borrow from cable theory the facts that
dendritic spikes may (or may not) be created by 
integrated synaptic input at some branchlets, then may (or may
not) propagate to neighboring branchlets, and annihilate upon
collision owing to refractoriness. Then, by employing a simplified
excitable model for each branchlet, but a realistic multicompartment
dendritic tree, we are able to focus on their collective behavior and
to cast the dynamics of the dendritic tuft into the framework of
extended excitable media, where both numerical and theoretical
approaches have been successfully
applied~\cite{Lindner04,Kinouchi06a,Wu07,Copelli02,Copelli05a,Copelli05b,Furtado06,Copelli07,Assis08,Ribeiro08a}.

Conventional wisdom in computational neuroscience is that in the limit
of a very large number of compartments the model would be physically
accurate. But in this same limit, conventional wisdom in statistical
physics (say, renormalization group arguments) tells us that
collective behaviors should be very weakly dependent on the detailed
modeling of the basic (compartmental) unit~\cite{Binney92}.
Macroscopic properties of extended media would rather depend more
strongly on dimensionality, network topology, symmetries, presence of
parameter randomness (disorder), noise, boundary conditions etc.
Therefore modeling should concentrate efforts on these more decisive
aspects, the use of simple excitable dynamics for the elementary
units being justified as a first approximation.

\paragraph*{Elementary dynamics.} 
 To account for the active nature of dendritic branchlets, each site
 is modelled as a simple discrete excitable element:
 $s_i(t)\in\{0,1,2\}$ denotes the state of site $i$ at time $t$
 (Fig.~1A). If the $i$-th branchlet is active ($s_i=1$), in the next
 time step it becomes refractory ($s_i=2$). The average refractory
 period is controlled by $p_\gamma$, which is the probability with
 which sites return to a quiescent state ($s_i=0$) again. Only
 quiescent sites can become active due to activation by neighboring
 compartments or by external (synaptic) inputs. Owing to the sometimes
 small density of ionic channels in dendrites, transmission of
 excitations from active to quiescent sites is modelled to occur with
 probability $p_\lambda$ per bond (see Fig.~1B). This means that, in
 the model, the propagation of excitation from branchlet to branchlet
 is not deterministic and may fail with probability $1-p_\lambda$.

\paragraph*{Dendritic topology.} 
We can think of an active dendritic tree as an excitable network in
which each site represents, for instance, an excitable dendritic
branchlet connected with two more distal sites and one more proximal
site (see Fig.~1B). That is, each branch at ``generation'' $g$ has a
mother branch from ``generation'' $g-1$ and gives rise to two daughter
branches at ``generation'' $g+1$ (i.e. each site ---
  except those at the borders --- has three neighbors). The most
distal generation will be called level $G$ and we will study tree
properties as a function of the branching order $G$. The single site
at $g=0$ would correspond to the primary dendrite which connects with
the neuron soma (see Fig.~1B). Notice that the number $N$ of
branchlets grows exponentially with the branching order $G$.

\paragraph*{External stimuli and branchlet activation.}
Each branchlet receives a large number of synapses, whose
post-synaptic potentials (excitatory and inhibitory) are
integrated. The final outcome of this complex integration (which we do
not model here) may or may not trigger a branchlet spike, which we
denote as our $s=1$ active state. As a first approximation, we assume
that this branchlet activation process (or crossing of the
excitability threshold) is Poisson with rate $h$, which somehow
reflects the average excess of excitation as compared to
inhibition. Thus, besides transmission from active neighbors, each
quiescent branchlet can independently become active with probability
$p_h \equiv 1-\exp(-h\delta t)$ per time step (see Fig.~1B), where
$\delta t=1$~ms is an arbitrary time step.

We assume that the activation processes of different branchlets are
independent of one another. Besides, we consider first the uniform
case where all branchlets have the same excitation rate $h$, which is
perhaps a reasonable assumption e.g. for mitral cells in the olfactory
system~\cite{Kosaka01}. We recognize that these are strong
simplifications, but the analysis of this case is essential as a first
step. Later we discuss non-homogeneous cases where $h$ depends on the
generation $g$ or, for each branchlet, is drawn from a normal
distribution.

\subsection*{Dendritic tree output as a response function} 
We define the apical activity $F$ as the number of excitations ($s=1$
states) produced at the $g=0$ site, averaged over a large time window
($10^4$ time steps and five realizations, unless otherwise stated). In
the following we will be interested in understanding the function
$F(h,N)$, which is somehow analogous to the neuron frequency versus
injected current $F(I)$ curves studied in the neuroscience
literature. We suppose that, in the absence of lateral inhibition, the
neuron firing frequency produced at the axonal trigger zone will be
proportional to the apical activity $F$, which is assumed by some
biophysical models~\cite{Migliore05} and supported by recent
experimental evidence in the {\it Drosophila\/} olfactory
system~\cite{Root2007}.

For readers familiar with statistical physics models we observe that
$F$ is the order parameter and 
$h$ is an external field that drives
the system to an active state with $F > 0$. Our model is an
out-of-equilibrium system with one absorbing
state~\cite{Marro99}. This means that, in the absence of external
drive ($h\to 0$) the dynamics eventually takes the system to a global
resting (quiescent) state $F =0$ from which it cannot escape without
further external stimulation. In biological terms, this simply means
that our dendritic tree will not show spontaneous dendritic spikes
without external synaptic input and any activity in the tree will
eventually die if $h$ is turned to zero.

% Results and Discussion can be combined.
\section*{Results}

\subsection*{Output dependence on arbor size $N$}
Dendritic trees are responsible for processing incoming stimuli which
impinge continuously on the many synaptic buttons spread on the
dendrites (a single olfactory mitral cell can have around $30,\!000$
synapses, whereas cerebellar Purkinje cells have around $200,\!000$
synapses). Of course these numbers vary also between individual cells
of the same type. So, we first consider a classical question asked
(and not clearly answered) in the literature: given a constant
activation $h$ in cells with different arbor sizes, will they fire at
very different levels~\cite{Spruston08}? The answer is not obvious
since they may have a huge difference of absolute number of synapses
and branchlets and we could have the prejudice that cells that have
more synapses should fire more easily (or at least need to implement
some homeostatic mechanism for controlling their firing rate).

The answer provided by our model is very interesting: for low
excitation rate $h$, the output $F(h,N)$ increases linearly with the
number $N$ of branchlets, so that having a large branched tree is
indeed important to amplify very weak signals (see Fig.~1C). In this
context there is a clear reason for a neuron to maintain a costly
number of branchlets. However, for moderate and high activation
levels, the activity $F(h,N)$ depends very weakly on $N$ (it grows
sub-logarithmically with $N$, see Fig.~1C for $N$, say, larger than
$5,000$). That is, in this regime the output $F$ reflects, in an
almost size-independent way, mostly the Poisson rate $h$, not the
absolute number of branchlets activated on the tree.

Large dendritic arbors therefore aid the detection of weak stimuli,
but for higher activation levels (i.e. higher imbalance between
excitatory and inhibitory signals) all the neurons code in a similar
way the activation rate $h$, irrespective of their arbor size. Note
that this ``size invariance'' is an intrinsic property of the
excitable tree, not based on any homeostatic regulatory
mechanism~\cite{RumseyAbbott,Turrigiano99}.  This sublogarithmic
dependence of $F$ on $N$ means that neurons function as reliable
transductors for the signal $h$: the specific number of branchlets,
developmental defects, or asymmetries of the dendritic tuft have only
a secondary effect in the global neuron functioning.

\subsection*{Output dependence on excitation rate $h$}

Given that the cell output depends weakly on $N$, now we turn our
attention to how $F(h,N)$ depends on the activation rate $h$. Note
that not much modeling work has been done on addressing the collective
activity of the dendritic tree subjected to extensive and distributed
synaptic input~\cite{PoiraziMel03a,PoiraziMel03b}, particularly as the
activation rate $h$ is varied. However, this is one of the simplest
questions one may ask regarding dendritic signal processing. In
particular, studies with models where the whole dendritic tree is
reduced to a small number of compartments (reduced compartmental
models~\cite{Herz06}) can hardly address this issue, since the complex
spatio-temporal information of the tree activation is lost by
definition.

As is well known, the average firing rate $F$ dependence on stimulus
rate $h$ of several cells has a saturating aspect like that of
Fig.~2A. Our cell presented a similar behavior (Fig.~2B), although of
course it is not the simple Hill function $F_{Hill}^{(m)}(h) =
F_{max}h^m/(h_0^m + h^m)$ usually employed to fit experimental
data. Indeed, for some values of axial transmission $p_\lambda$, we
saw an unexpected double-sigmoid behavior (see below).

\paragraph*{Tree dynamic range.} The dynamic range $\Delta$ of the response
function follows a standard definition:
\begin{equation}
\label{delta}
\Delta = 10\log\left(\frac{h_{90}}{h_{10}}\right)\; ,
\end{equation}
where $h_{90}$ ($h_{10}$) is the stimulus value for which the response
reaches $90\%$ $(10\%)$ of its maximum range ($F_{90}$ and $F_{10}$
respectively). As exemplified in Fig.~2A with a Hill function,
$\Delta$ amounts to the range of stimulus intensities (measured in dB)
which can be appropriately coded by $F$, discarding stimuli which are
either so weak as to be hidden by noise or self-sustained activity of
the system ($<F_{10}$) or so strong that response is in practice
non-invertible owing to saturation ($>F_{90}$). It is a
straightforward exercise to show that for a general Hill function with
exponent $m$ we have $\Delta = (10 \log 81)/m \approx 19/m$~ dB. This
simple analytical result reinforces the fact that the exponent $m$
governing the low-stimulus regime is determinant for the dynamic
range.

Figure~2B shows how the response curve changes with the coupling
$p_\lambda$ between dendritic patches. For $p_\lambda=0$ (lowest
curve) each dendritic patch is an isolated excitable element, so
activity does not spread in the tree and the response function is a
linear-saturating curve with a small dynamic range ($\simeq 16$~dB,
see Fig.~3). As $p_\lambda$ increases, more signals are transmitted to
the apical site. This amounts to an amplifying mechanism whose
efficiency increases with $p_\lambda$, as depicted in Fig.~2B.

Amplification, however, is highly nonlinear. Note that a dendritic
spike dies after some time either by propagation failure
or, more importantly, because it is annihilated upon collision
with other excitable pulses or with the tree boundaries (the $g=G$
distal branches). Since the likelihood of these collisions increases
with the stimulus intensity, amplification is stronger for weak
stimuli and weaker for strong stimuli (in particular, for very strong
stimulus every excitable element reaches its maximum activity --
limited by refractoriness -- and coupling is almost irrelevant). As a
consequence, sensitivity {\em and\/} dynamic range are concurrently
enhanced with increasing coupling, as illustrated in Fig.~3.

We emphasize that the above reasoning relies on very general and
robust properties of excitable media: any detailed compartmental
biophysical model of an active dendritic arbor will present similar
results. Two features, however, strike as particularities of a tree
topology: 1) the dynamic range attains extremely large values (see
Fig.~3) and 2) the response functions can become double-sigmoids, due
to interaction with dendritic backspikes, as depicted in the upper
curves of Fig.~2B and discussed below.

%\subsection*{Subsection 1}

%\subsection*{Subsection 2}

\section*{Discussion}

A possible critique to our modeling approach is that it lacks
biological realism. We notice that this is only true at the level of
the biophysical dynamics of each compartment, but we believe that the
idealization of such compartment as a generic excitable element (a
cyclic automaton) is immaterial. This has already been demonstrated in
studies of the dynamic range of networks composed by cellular
automata, non-linear discrete time maps, nonlinear differential
equations and conductance-based models (Hodgkin-Huxley
compartments)~\cite{Copelli05a, Ribeiro08a}, as well as in a
biophysically detailed model of the vertebrate
retina~\cite{PublioTese}.

Our model has realistic biological aspects not reproduced by most works in
computational neuroscience with detailed biophysics: 
\begin{itemize}
\item a tree topology with a very large number ($\approx 10^5$) of compartments
(branchlets); 
\item a proportionally large number of synaptic inputs;
\item distributed activation along the whole tree instead of
artificial injected currents applied at particular
points. 
\end{itemize}

Notwithstanding the fact that artificial input protocols like punctual
current injection are useful for comparison with
experimental measurements~\cite{DeSchutter94}, we believe that 
spatio-temporal Poisson
activation is a step toward a more realistic modeling of the
dendritic arbor dynamics under natural
circumstances~\cite{RumseyAbbott,PoiraziMel03a,PoiraziMel03b}.

\subsection*{The dynamic range problem}
As can be viewed in Fig.~3, large active dendritic trees perform
strong signal compression, which is the ability of coding many orders
of magnitude of stimulus intensity through only one decade of output
frequency. This question is particularly important in sensory
processing, where {\em many\/} orders of magnitude of stimulus
intensity are present. Interestingly, olfactory glomeruli, constituted
primarily by large active dendrites of mitral cells in vertebrates and
dendrites of principal cells in insects have large dynamic
range~\cite{Friedrich97,Wachowiak01,Bhandawat07}. We conjecture that a
similar situation occurs in the problem of fine motor control and
sensory-motor integration in the cerebellum~\cite{Gao96}, which also
involves the necessity of handling sensory-motor feedback signals
varying by orders of magnitude. In correspondence to our hypothesis,
Purkinje cells, which are involved in these tasks, have indeed
enormous active dendritic arbors~\cite{RothHausser01,Shepherd}.

Previous
work~\cite{Kinouchi06a,Wu07,Copelli02,Copelli05a,Copelli05b,Furtado06,Copelli07,Assis08,Ribeiro08a}
has shown that the non-linear summation of spikes enhance the dynamic
range of excitable media. 
The tree topology, however, has not been studied in these works.
Surprisingly, we found that its performance is largely superior to the
others. This motivates the proposal, first made here (to
the best of our knowledge), that the main functional role of active
dendrites is to enlarge the cell dynamic range. 

As a particular application, we discuss now the case of the dynamic
range of olfactory glomeruli.  Recent results for second-order
projection neurons of the {\em Drosophila melanogaster\/} antennal
lobe clearly exhibit strong weak-stimulus amplification and enhanced
dynamic range as compared to olfactory receptor neurons
(ORNs)~\cite{Bhandawat07}.  To account for this observation, we can
interpret our model as representing a {\it Drosophila\/} principal
cell (analogous to a mitral cell) inside the glomerulus.  Also, the
signal propagation from ORN axons to principal cell dendrites and the
proportionality between apical activity and somatic firing measured by
Root {\em et al.} in the {\it Drosophila\/} is compatible with our
identification of $F$ with the somatic neuron response~\cite{Root2007}
in this particular case.  These authors show that it is mainly the ORN
activity that drives the projection neuron firing rate, the isolated
effect of synapses from interneurons (excitatory and inhibitory) being
not sufficient to induce spikes and having mostly a modulatory role.

Of course, in the case of other biological systems like the mammalian
olfactory bulb (where strong lateral inhibition occurs) or pyramidal
cells, the identification of $F(h)$ with the somatic firing rate is
problematic, but we claim that the model is still useful for
understanding of the large dynamic range (as measured by Calcium
fluorescence) observed in the neuronal
tuft~\cite{Friedrich97,Wachowiak01}.

It is important to notice that large dynamic ranges as observed here
means that the output varies slowly with the input. Therefore, if
experiments are done over only one or two orders of magnitude of
stimulus intensity (10-20 dB), the observed effect could be confounded
with an almost constant response.  This may be an alternative
explanation for the \textit{concentration invariance} property
observed in olfactory processing~\cite{Wilson06}.

\subsection*{Weak dependence of activity on branchlet number} 
Another important prediction of our model is that dendritic size (and
the respective number of branchlets and synapses) has a weak effect on
the apical activation, being important mostly in the small excitation
regime. It is mainly the branchlet activation rate $h$, not the total
number of branchlets, that controls the apical rate $F$.  This is a
desirable robustness property since there is a high variability of
dendritic size and spine density within a neuron population and along
time in the same neuron.

Whichever function one wishes to assign to active dendrites, it must
be fault tolerant in relation to gross dendrite morphology, branchlet
excitability and synaptic density, which vary with age and time: for
example, 30\% of spine surface retracts in hippocampal neurons over
the rat estrous cycle~\cite{Woolley90}.  Due to the sublogarithmic
dependence of $F(h,N)$ on $N$ (see also the Model Robustness section),
our model demonstrates that such gross independence from branchlet
number, detailed branchlet dynamics, dendritic axial conductance and
tree morphology is possible, and that enhancement of dynamic range is
one of the most visible properties of these excitable trees.

\subsection*{Response functions with double sigmoids}

Double-sigmoid response functions have been reported recently for
retinal ganglion cells of the mouse~\cite{Deans02}. This unusual shape
contrasts with the standard Hill fitting function.  One wonders
whether the habit of fitting Hill functions to data could have
prevented further double-sigmoid curves from having been reported in
the literature.

It is very interesting that such double sigmoid behavior is a
distinctive feature of our model in a certain range of parameter
space. Can we interpret the findings on retinal ganglionar cells in
terms of our simplified model of dendritic response? Ganglionar cells
have dendritic arbors but their size is small compared to, say, mitral
cells or our typical model with branching order around $G = 10$.

However, in a structural analogy between the visual and olfactory
systems, Shepherd proposed that some ganglionar cells are the retinal
equivalent of mitral cells~\cite{Shepherd}. Here we pursue this
analogy and suggest that the ganglionar dendritic arbor plus the
retinal cells connected to it by gap junctions (electrical synapses)
can be viewed as an extended active tree similar to the one studied
here, with a large effective $G$.

We show in Fig.~2D that an appropriate choice of the model parameters
can lead to a response function which fits the experimental data. Of
course, the quantitative fit, although good, is not the important
message, but the qualitative one: that double-sigmoid response
functions can appear solely due to the tree topology, without invoking
any secondary activation processes or complicated mechanisms to
produce the unusual shape.

What is the physical origin of the double sigmoids in our model?  We
believe that it is related to the two different modes of activation of
the apical site. The first one is the direct excitation due to its
local $h$ rate.  This direct excitation, if large, drives the system
to its maximum firing rate, which scales with the inverse of its
refractory period. This mechanism would be responsible for the
saturation in the right side of $F(h)$ (region of large $h$), see
Fig.~2B.

But the apical site also receives signals from its extended dendritic
tree, which is very sensitive to small activity (extending the $F(h)$
curve to the small $h$ regime). However, it is plausible that the tree
excitability saturates for a smaller frequency, due to the complicated
interations between the spikes in the tree. So, we conjecture that the
first sigmoid represents a bottleneck effect related to saturation in
the flux of the activity along the subtrees connected to the apical
site.  Indeed, this is compatible with the observation that if we
disconnect the apical site from the dendritic tree ($p_\lambda =0$,
Fig.~\ref{fig:response}B), the double sigmoidal behavior disappears
and only the second (large $h$) sigmoid is maintained.  Of course, a
more detailed analysis of the origin of the first sigmoid is needed.

%We also observed curves with three sigmoids (see
%Fig.~\ref{fig:quenched}C), but postpone the discussion of these
%results to future works.

We also observed curves with three sigmoids (see Model Robustness 
section), but postpone the discussion of these results to future works. 
We only note here that the intermediate-$h$ 
plateau in these curves could also be related to the concentration 
invariance reported for olfactory systems~\cite{Wilson06}.

\subsection*{Screening resonance} 
As can be seen in Fig.~2B, some response curves in our model can
present an unusual shape, with curves for higher probability of axial
transmission $p_\lambda$ falling below curves for lower
$p_\lambda$. How can more efficient trees present a response below
less efficient ones for the same $h$ level?

This question can be answered by looking at Fig.~4A, where we plot a
family of curves $F(p_\lambda)$ for fixed $h$. For some (intermediate)
values of $h$, this curve is non-monotonic, suggesting a kind of
resonance through which activity in the primary dendrite is maximized
for an optimal coupling among sites all over the tree. Why is this so?

Note that, on the one hand, for low enough $p_\lambda$, excitations
created in distal sites may not arrive at the primary site due to
propagation failure. For too strong coupling, on the other hand, {\em
  the topology\/} of the tree leads to a dynamic screening of the
primary dendrite: backward propagation of activity (backspikes)
effectively can block forward propagation of incoming signals, as
shown in Fig.~4B. Activity $F$ is therefore maximized at some
intermediate value of coupling. We called this phenomenon ``screening
resonance''.  That such screening resonance indeed depends on
backspikes is confirmed by an asymmetrical propagation variant of the
model (see Model Robustness section). As backpropagation goes to zero,
the crossing between $F(h)$ curves disappears (Figs.~4C and~5C).

The transmission probability $p_\lambda$ accounts for the joint
effects of membrane axial conductance and density of regenerative
ionic channels (Na$^+$, Ca$^{2+}$, NMDA etc). A possible experiment to
test whether this screening resonance indeed exits could involve the
manipulation of the density (or efficiency) of those channels in the
dendritic tuft: the model predicts that more excitable trees may
present lower activity than less excitable ones due to resonant
annihilation of dendritic spikes.

\subsection*{Testing dendritic spike annihilation} 

As discussed above, annihilation due to collision of dendritic spikes
is the central mechanism in our model behind both the dynamic range
enhancement (by preventing the tree response to be proportional to the
rate $h$) and the screening resonance phenomenon (by blocking
forward-propagating dendritic spikes with backward-propagating ones).

With rare exceptions~\cite{RumseyAbbott,RoyerMiller07}, the fact that
nonlinear summation often implies spike annihilation has been somewhat
underrated in the literature. Recent simulations with biophysical
compartments show the propagation and collision of dendritic
spikes~\cite{RumseyAbbott,RoyerMiller07}.  To fully evaluate our
ideas, one should examine better this phenomenon in \textit{in vitro}
dendrites.  The computational results suggest the following simple
experimental tests:

\begin{enumerate}
\item After the simultaneous creation, by two electrodes, of counter
  propagating spikes on a long apical dendrite, no spike should be
  detected in either electrode due to spike annihilation in the space
  between them.
\item Blocking of active channels should reduce the dynamic range of
  cells with large dendritic arbors, but the effect would be less
  important in the case of cells with small dendrites (see
  Fig.~\ref{fig:delta}).  
\item After the simultaneous creation of spikes in two dendritic sites
  on the same subtree, the neuron output should be almost the same as
  that obtained with injection at a single point (due to spike
  collision at some branchlet of the subtree). The final output is not
  the (linear) sum of EPSPs but rather the tree functions as an OR
  gate if the injected currents are simultaneous and located at a
  similar level $g$.
\end{enumerate}

One consequence of spike annihilation is that under moderate
stimulation backspikes will fail to reach more distal branches, owing
to collisions with forward-propagating dendritic spikes and/or
refractory branches~\cite{RumseyAbbott,GoldingSpruston}. 
Indeed, we have observed this phenomenon in our model.

This is compatible with recent observations that backspikes are
strongly attenuated in the presence of synaptic input in medial
superior olive principal neurons~\cite{Scott2007}.  So, the use of
somatic backspikes as a backpropagating signal under massive synaptic
input seems to be problematic. Somatic backspikes show up naturally in
excitable trees but plays no functional role here. 

We conjecture that somatic backspikes may be epiphenomena or perhaps,
if they have a functional role in learning processes, it is a recent
evolutionary exaptation from previous robust functions like signal
amplification by dendritic spikes.  This can be tested: our model
predicts that active dendrites will be found even in neurons without
any plasticity or learning phenomena.

\subsection*{Relation to Psychophysics}

Our modeling approach also suggests a microscopic (neural) basis for
Stevens law of psychophysics~\cite{Stevens,Augustin08}, which states
that the perception $F$ of stimulus intensity $h$ grows as a power law
$F \sim h^m$.  In a previous work with disordered
networks~\cite{Kinouchi06a}, by assuming a linear relationship between
psychophysical perception and the network activity, we have found a
Stevens-like exponent for the input-output function of excitable media
with value $m = 0.5$. For planar networks we found $m \approx
0.3$~\cite{Assis08}.  Here we found for the dendritic tree
architecture that the Stevens exponent is very small ($m \approx 0.2$
or even $0.1$ for large trees with $G>20$), which means that the
response function could be confounded with a logarithmic
(Weber-Fechner) law~\cite{Stevens}. Of course, the macroscopic
psychophysical law would be a convolution of all these non-linear
transfer functions between the sensory periphery and the final
processing (psychological) stage.

What our model shows is that any excitable medium naturally presents a
nonlinear input-output response with exponent $m < 1$, that is, large
dynamic range, and that perceptual ``psychophysical laws'' could be a
very early phenomenon in evolution. A simple precondition is that the
sensory network should have an excitable spatially extended dynamics,
like the one already found in bacterial chemotaxis channel networks,
for example \cite{Bray98,Barkai98}.

\subsection*{Model robustness} 

In our model, variable branchlet diameter and size is described by a
spatial dependence and disorder in $p_\lambda$. We do not expect the
results concerning the dynamic range to change qualitatively with this
type of generalization. The same model robustness appears for changes
in the refractory time and the use of continuous dynamical variables
(maps or differential equations). This latter property has already
been demonstrated in multilevel modeling studies which used cellular
automata and nonlinear differential equations to describe the neuronal
excitable elements~\cite{Copelli05a, Ribeiro08a, PublioTese}.  We now
explicitly show the results for three variants of the model in order
to address the robustness of the dynamic range enhancement.

\paragraph*{Model I: Propagation Asymmetry.}

First, we consider the possibility that backward transmission of
excitation is less likely to occur than in the forward direction (as
suggested by impedance matching arguments). For that purpose, we keep
$p_\lambda$ for denoting the probability of transmission in the
forward direction and let $\beta p_\lambda$ be the probability of
transmission in the backward direction, with $0\leq \beta \leq 1$.

For $\beta=1$, a given active branchlet at level $g$ excites a
quiescent daughter at $g+1$ and its quiescent mother at $g-1$ with the
same probability (this corresponds to the results presented so far).
At the other extreme, for $\beta=0$, backpropagating dendritic spikes
do not occur at all.

In Fig.~5A we show the $F(h)$ curves for varying $\beta$, which
present pronounced double-sigmoid behavior, suggesting that backspikes
regularize the first sigmoidal saturation. In Fig.~5B we see that the
dynamic range $\Delta(p_\lambda)$ curves are almost the same for
different values of $\beta$ (note that the differences among the
$F(h)$ curves occur for intermediate values of $h$, so the $h_{10}$
and $h_{90}$ points remain essentially the same).  Therefore, a
possible functional role for backspikes could be to regularize the
response curve, since the first saturation of a double-sigmoid
corresponds to poor coding. Of course, this presumed functional role
is only speculative, but is a new suggestion provided by our model.

Another result of this variant of the model confirms that backspikes
are indeed responsible for the crossing of the response curves $F(h)$
for different values of $p_\lambda$ (see Fig.~2B). As depicted in
Fig.~4C, the screening resonance phenomenon only occurs for $\beta$
sufficiently close to one. 

We also examined the effect of varying the coupling between branchlets
($p_\lambda$, Fig.~5C) and their refractory period ($p_\gamma$,
Fig.~5D) in the asymmetric propagation model with $\beta = 0.5$.  We
see some new phenomena like a non-monotonic dependence of $\Delta$ on
the coupling $p_\lambda$ (Fig.~5D), which only occurs for $\beta<1$.

\paragraph*{Model II: Non-homogeneous branchlet activation.}

Since the rate $h$ reflects the imbalance between synaptic excitation
and inhibition at the branchlets, and since different branchlets may
receive a different number of synapses, a step torward more realistic
modeling would involve a branchlet-dependent $h$. We investigate the
effects of a kind of non-homogeneity present in several neurons, where
synaptic density or excitability tends to be larger in more distal
branchlets.

A simple model that incorporates this spatial dependence is $h(g) =
h_0\exp(ag)$. For $a=0$ we recover the homogeneous model, while for
$a>0$ the rate of dendritic spike creation increases with the distance
from the soma (note in particular that for $a \ll G^{-1}$ this
increase is approximately linear).  The use of an exponential model is
motivated by the fact that it is an extreme one: any polynomial
dependence model lies between the uniform and the exponential case. If
the exponential model does not produce qualitative changes, then
polynomial models hardly will do.

In Fig. 6A we show the response curves $F(h_0)$ for several values of
the parameter $a$. In Fig.~6B we show that the enhancement of dynamic
range obtained from the response curves $F(h_0)$ (for different values
of $a$) is robust. Surprisingly, the signal amplification and dynamic
range is indeed much more efficient than the homogeneous case $a=0$,
attaining $80$~dB (notwithstanding the poor coding for intermediate
values of $h$, where the size of the plateau increases with $a$ and
$p_\lambda$). This result suggests that this case of peripheral
branchlets being more excitable could optimize the signal processing,
specially for neurons with poor propagation of dendritic spikes (small
$p_\lambda$, see Fig.~6B).

\paragraph{Model III: Disordered branchlet activation rate}

In the previous variant, all branchlets in the same generation $g$
have the same activation rate $h(g)$.  Now, we study a disordered $h$
model, where each branchlet $i = 1,\ldots,N$ is initially assigned a
rate $h_i=u_i \kappa h_0 + h_0$. The parameter $\kappa$ is fixed for
each curve and $u_i$ is drawn from a Gaussian distribution with zero
mean and unit variance, and is kept constant throughout each run. Note
that $\kappa$ corresponds to the coefficient of variation $\sigma/h_0$
of the distribution $P(h)$, where $\sigma$ is the standard
deviation. Since the Poisson excitation rate $h$ must be positive, we
set $h_i = 0$ if we some branchlet gets a $h_i < 0$ from the Gaussian
distribution.
 
The response curves $F(h_0)$ for different values of $\kappa$ are
shown in Fig.~6C. The enhancement of dynamic range is essentially
unchanged even under strong variability ($\kappa=1$) of branchlet
activation rate (Fig. 6D).

\subsection*{Conclusions and perspectives}

Several detailed biophysical models of dendritic trees have already
been presented in the literature, but we are not aware of studies
confirming the enlargement of the dynamic range by active dendrites in
such arbors. To see this effect, it is necessary that such models
incorporate inputs distributed along the full dendritic
tree, and that the branchlet activation rate 
be varied by orders of magnitude. 

We believe that biophysical multi-compartmental models (with a large
number of branchlets) seeking to probe the robustness of our results
would be most welcome. In particular, they would be able to address
the effect of post-synaptic potentials (PSPs, both excitatory and
inhibitory) which manage to generate somatic -- but not dendritic --
spikes despite the presence of active channels in the dendrites (a
phenomenon which might be artificially adapted to our model, but for
which biophysical models are better equipped). Also, the modulatory
effect of such subthreshold PSPs and other passive phenomena are
better studied in biophysical simulations.
  
Other future tasks will be the study of the dendritic response due to
non-Poisson input distributions (say, $1/f^\beta$ noise), correlated
input on the arbor, time-dependent inputs, asymmetric dendritic trees
etc. We believe that new signal processing features may appear, but
the dynamic range enlargement and sensitivity enhancement by active
dynamics will continue to be present.

Why do neurons have active channels in extensive dendritic trees?  Our
proposal is that active large dendrites are able to detect and amplify
very weak signals and, at the same time, saturate slowly for stronger
tree activity.  This universal ``dynamic range'' problem, related to
the trade-off between sensitivity and saturation of signal processing,
is important both for individual neurons, large neural networks, whole
sensory organs and organisms. We conjecture that the large dynamic
ranges found in neurons with active dendritic arbors could even help
to explain macroscopic psychophysical laws, providing a neural account
for the century old findings of Fechner, Weber and
Stevens~\cite{Stevens,Kinouchi06a,Chialvo06}.

% Do NOT remove this, even if you are not including acknowledgments
\section*{Acknowledgments}

The authors acknowledge financial support from Brazilian agencies
Conselho Nacional de Desenvolvimento Científico e Tecnológico (CNPq),
CAPES and FACEPE, as well as special programs PRONEX and Instituto
Nacional de Ciência e Tecnologia em Interfaces Cérebro-Máquina
(INCEMAQ). LLG was also supported by the European Commission Project
GABA (FP6-NEST Contract 043309), and the MEC (Spain) and Feder under project FIS2007-60327 (FISICOS). The funders had no role in study
design, data collection and analysis, decision to publish, or
preparation of the manuscript. 
The authors are also grateful to A. C. Roque, M. A. P. Idiart,
G. L. Vasconcelos and R. Dickman for discussions, as well as two
anonymous referees for their suggestions of the variants of the model. 

%\section*{References}
% The bibtex filename
\bibliography{copelli}

\section*{Figure Legends}

\begin{figure}[!ht]
\begin{center}
\includegraphics[angle=0,width=0.6\textwidth]{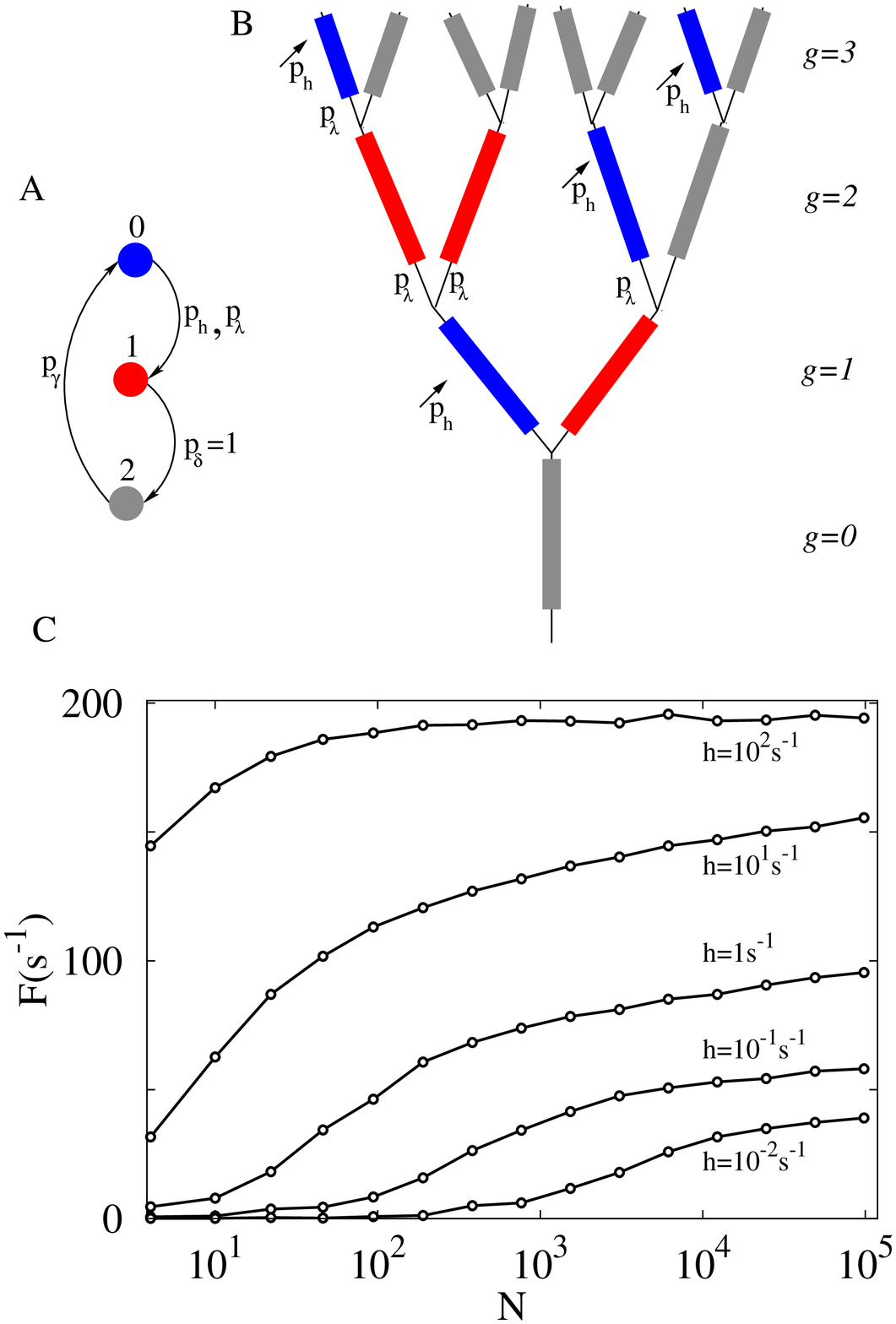}
\end{center}
\caption{ 
{\bf Morphology and dynamics of the model.} (A) Definition of
dynamical states: each dendritic branchlet can be in one of three
states (represented by circles): quiescent (blue), active (red) or
refractory (grey). A quiescent state becomes active due to integrated
synaptic input (with probability $p_h$) or transmission from an active
neighbor (with probability $p_\lambda$, also called the coupling
parameter). The active state has a fixed duration, changing to the
refractory state after a single time step ($p_\delta = 1$).  The
refractory state returns to the quiescent state with probability
$p_\gamma$ ($= 0.5$ unless otherwise stated). (B) Example of an active
dendritic tree with $G=3$: branchlets connected in a binary tree
topology. The probability that activity in one branchlet activates its
neighbour is $p_\lambda$ (if the neighbor is in a quiescent
state). (C) Apical activity $F(N)$ as a function of the number $N$ of
dendritic branchlets. Due to integrated synaptic input, each branchlet
becomes excited with a probability distribution modeled as an
independent Poisson process with rate $h$, as well as deterministic
propagation from active neighbors ($p_\lambda=1$). From bottom to top:
$h = 0.01, 0.1, 1.0, 10, 100$ activations per second at each
branchlet.}
\label{fig:dendrite} 
\end{figure}

\begin{figure}[!hbt] 
\begin{center}
\includegraphics[width=0.99\textwidth]{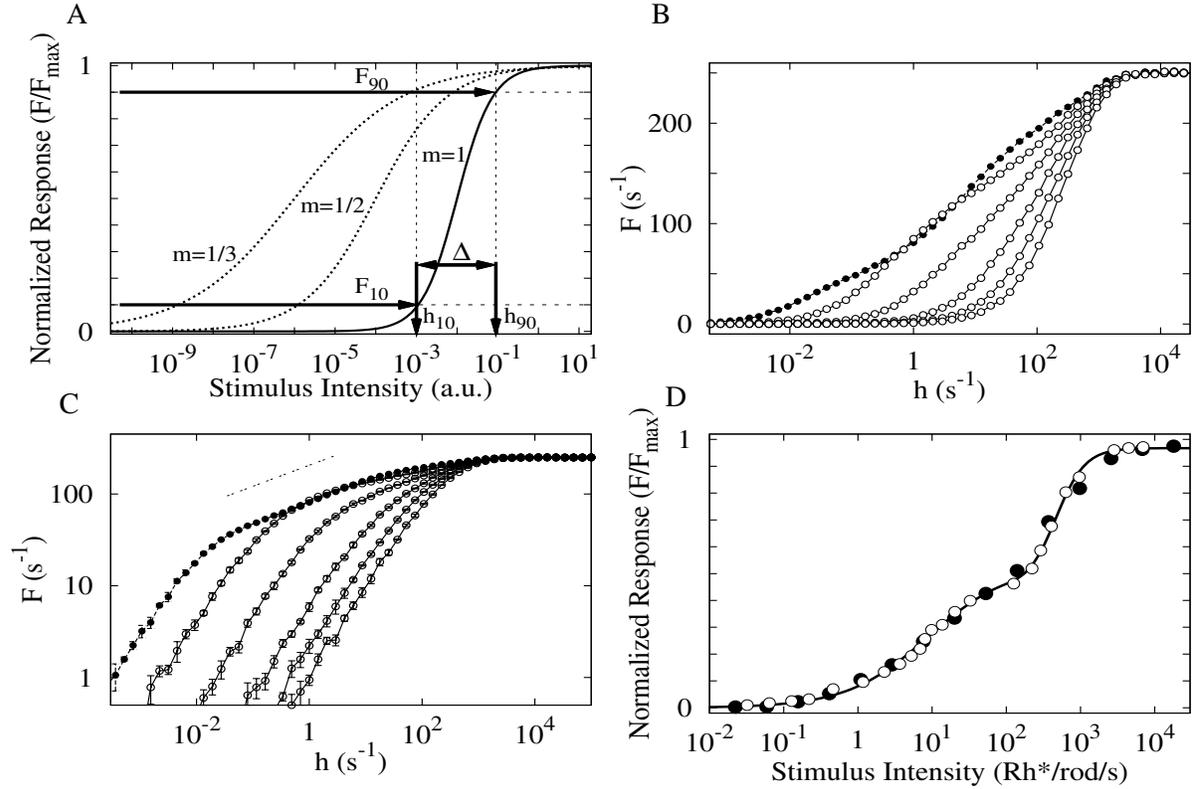}
\end{center}
\caption{ 
{\bf Response functions $F(h)$.} (A) Response functions exemplified by
normalized Hill functions $F/F_{max} = h^m/(C + h^m)$ with different
Hill exponents $m$. Relevant parameters for calculating the dynamic
range are exemplified for $m=1$, in which case $\Delta = 10\log 81
\simeq 19$~dB (see Eq.~\ref{delta} and text for details). (B) Family
of response curves $F(h)$ for $G=10$. From bottom to top, open symbols
represent $p_\lambda=0, 0.2, \ldots, 0.8$, whereas closed symbols
represent a deterministic transmission of activity ($p_\lambda=1$)
between dendritic branchlets. For $p_\lambda > 0.5$ extra inflection
points appear, giving rise to double sigmoid functions. (C) Study of
the response exponent. Same curves as (B), but in double logarithmic
scale. Notice the emergence of a non trivial and very small exponent
$m$ ($\approx 0.2$, thin dashed line) when reliability of dendritic
spike propagation ($p_{\lambda}$) increases. Notice also that, for
small input, spikes seldom colide: the output frequency $F$ is thus
proportional ($m = 1$) to the rate of branchlet activation (which
creates the spikes). (D) Double sigmoid experimental response curve of
retinal ganglion cells extracted from Ref.~\cite{Deans02} (open
symbols) compared to simulation results (closed symbols) for $G=15$
and $p_\lambda = 0.58$. To scale the model variable $h$ (ms$^{-1}$) to
the experimental stimulus intensity $I$ (Rh*/rod/s), we have employed
$h = 0.42 I$. The solid curve is a fit of two different Hill functions
joined together at $110$~Rh*/rod/s.}
\label{fig:response} 
\end{figure} 

\begin{figure}[!htb] 
\begin{center}
\includegraphics[angle=0,width=0.9\textwidth]{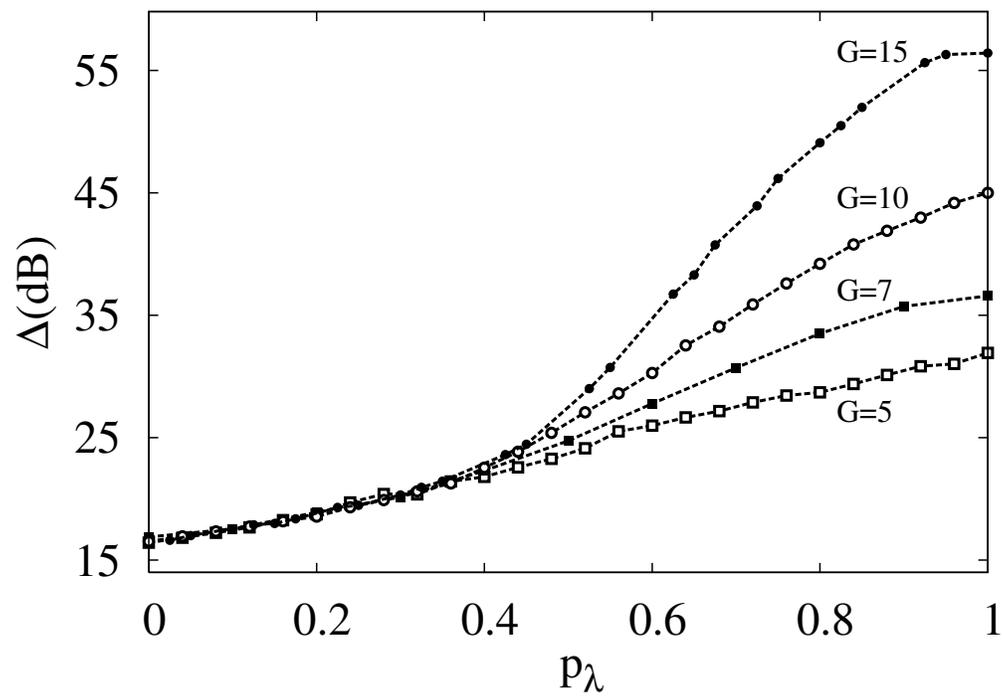}
\end{center}
\caption{ 
{\bf Enhancing the dynamic range.} Dendritic trees perform non-linear
input-output transformations such that the capacity to distinguish
between different stimulus intensities, measured by the dynamic range
($\Delta$) increases monotonically with coupling $p_{\lambda}$ and the
tree size ($G$). The tree topology can produce very large dynamic
ranges (above 50~dB). }
\label{fig:delta} 
\end{figure}

\begin{figure} 
\begin{center} 
\includegraphics[angle=0,width=0.95\textwidth]{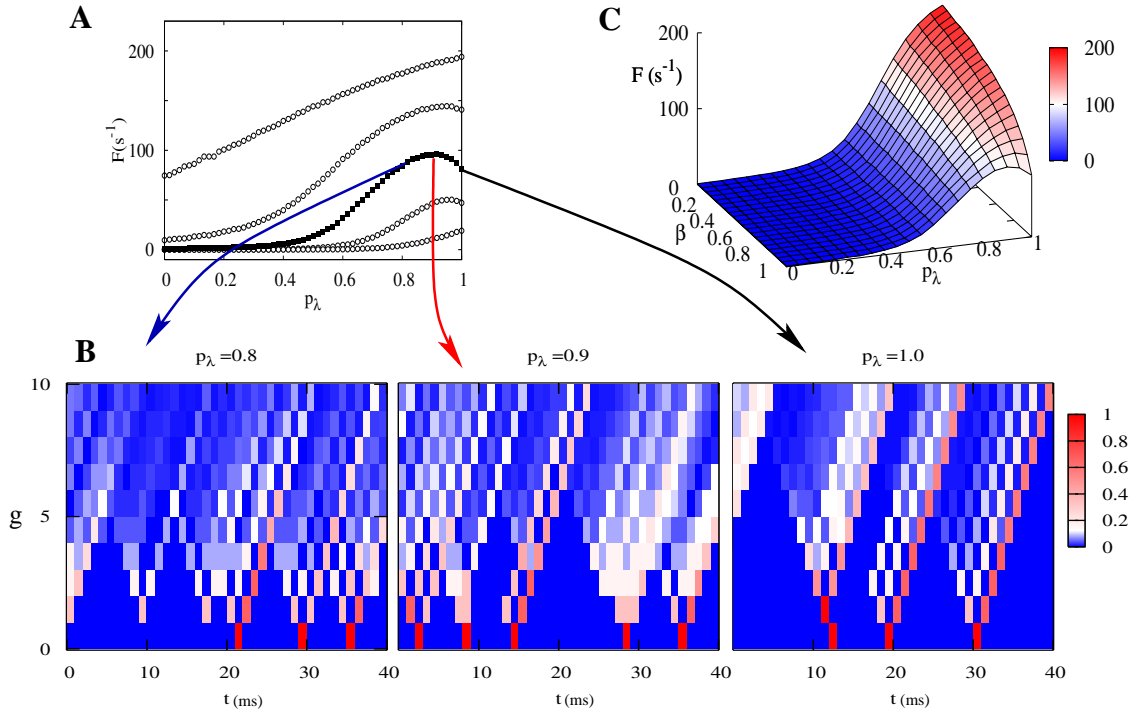} 
\end{center} 
\caption{ 
{\bf Screening resonance.} Depending on the rate $h$, a maximum on the
neuronal apical activity may be observed at an intermediate coupling
value ($p_{\lambda}$). Propagation of forward signals fails to
effectively induce neuronal apical activity for higher values of
coupling due to backpropagating activity in a certain range $h$ (here,
the retropropagation ratio is $\beta = 1$). (A) The non-monotonous
behavior in the mean output frequency $F$ at the primary dendrite as a
function of the coupling $p_\lambda$ among sites (closed symbols).
From bottom to top, $h = 0.01, 0.1, 1.0, 10, 100$ activations per
second per branchlet. (B) Density of active branchlets at generation
$g$ vs. time for $G=10$ and $h=1$~s$^{-1}$. Notice that in this short
(40 ms) sample, apical activity was higher for $p_\lambda = 0.9$ (5
activations) than for $p_\lambda=0.8$ or $p_\lambda=1$ (3 activations
each). The backpropagating signal for $p_\lambda=1$ prevents distal
activity from reaching the apical branchlet. (C) $F$ as a function of
backpropagation ratio $\beta$ and coupling probability $p_\lambda$,
for fixed $h=1$~s$^{-1}$: the screening resonance disappears in the
absence of backspikes (low values of $\beta$).}
\end{figure}

\begin{figure} 
\begin{center} 
\includegraphics[angle=0,width=0.95\textwidth]{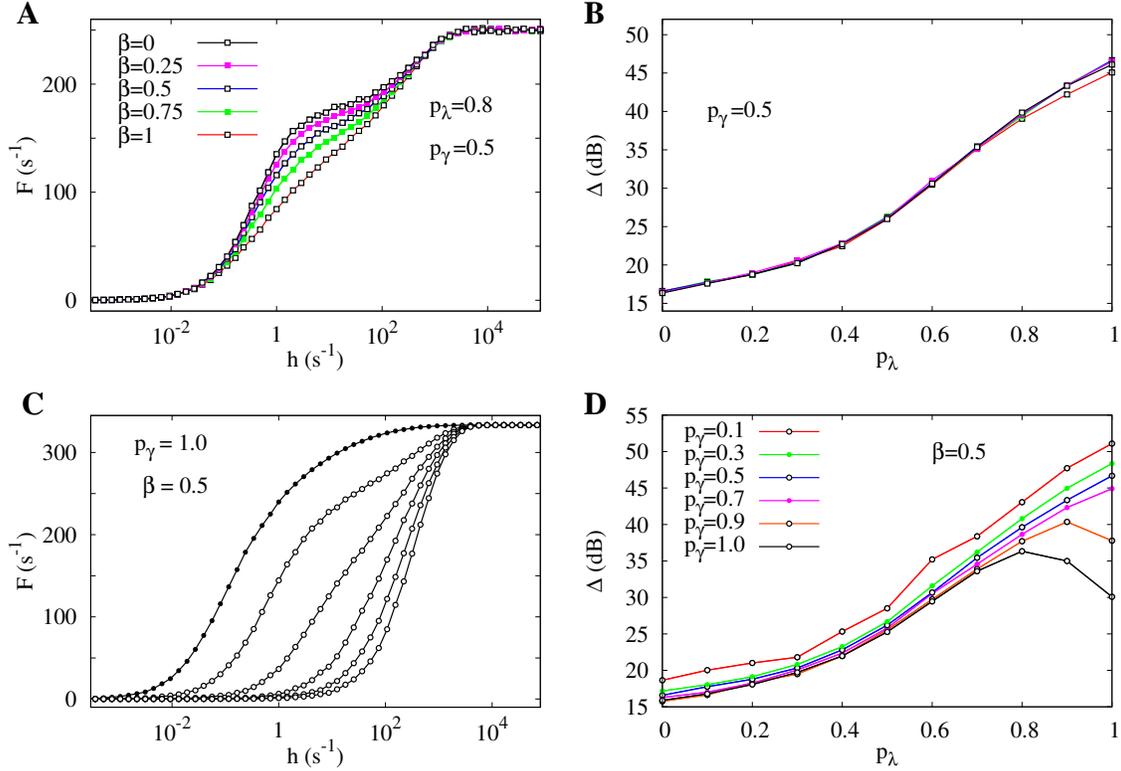} 
\end{center} 
\caption{ 
{\bf Effect of asymmetric propagation.} (A) Response functions for
different values of backpropagation ratio $\beta$ for fixed values of
transmission probability $p_\lambda=0.8$ and recovery probability
$p_\gamma=0.5$ (which controls the refractory period). It shows a
specific shape dependence with more visible double sigmoid behavior
for less backspike activity (lower values of $\beta$). (B) Dynamic
range of the response functions shown in panel (A). Although the
response functions $F(h)$ have different shapes, their dynamic ranges
remain pretty much unaltered since the region in which the response
functions differ is located in between the range of $h_{10}$ and
$h_{90}$ (see definition of dynamic range in Fig. 2A).  (C) A family
of response functions for deterministic refractory period
($p_\gamma=1.0$) and asymmetric propagation ($\beta=0.5$). Similarly
to Fig.~2B, from bottom to top open symbols represent
$p_\lambda=0,0.2,...,0.8$, and filled circles represent the case of
$p_\lambda=1$. The model presents a wide variety of response function
shapes. The filled symbols present a dynamic range smaller than for
$p_\lambda=0.8$, which is a rare example of non-monotonicity of the
$\Delta(p_\lambda)$ dependence. It occurs because the gain in
sensitivity (of $h_{10}$) when $p_\lambda$ increases from $0.8$ to $1$
is less than what is lost due to an early saturation (of
$h_{90}$). (D) Dynamic range for different values of refractory
period. The black curve displays the non-monotonicity explained in
panel C).  Besides this feature (which occurs only for $\beta<1.0$)
the dynamic range does not present qualitative changes compared to the
standard symmetric model of Fig.~3.}
\label{fig:beta} 
\end{figure}

\begin{figure} 
\begin{center} 
\includegraphics[angle=0,width=0.95\textwidth]{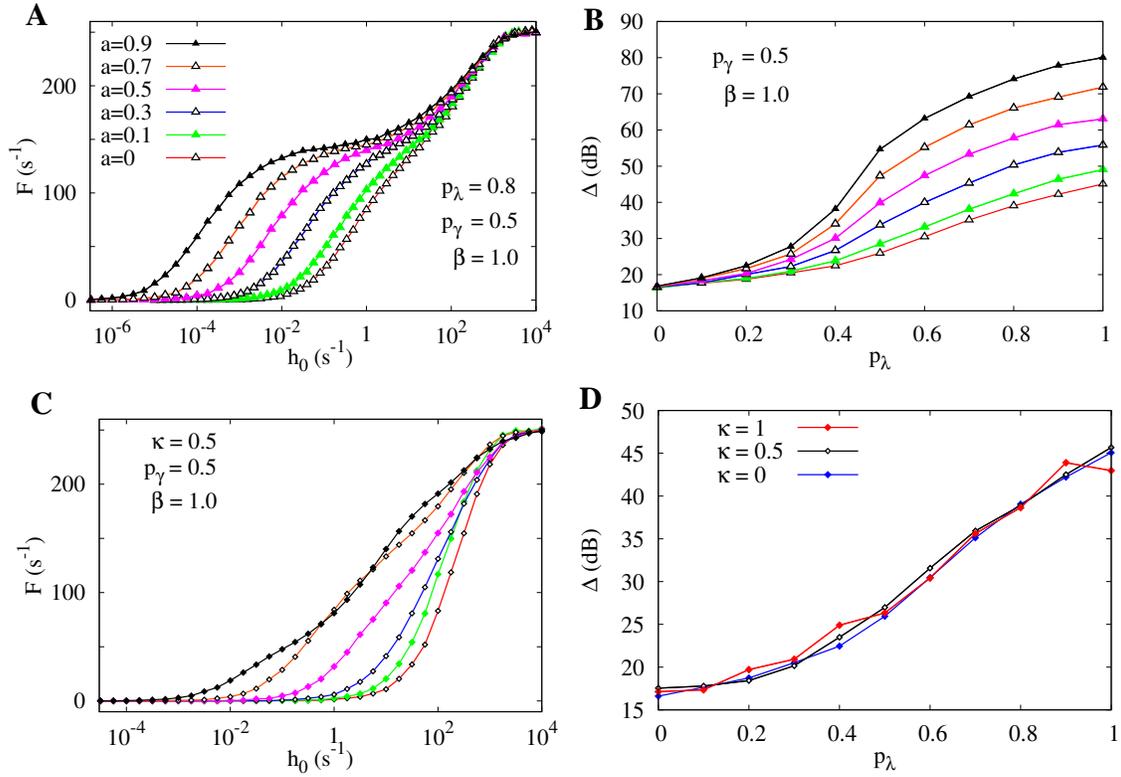} 
\end{center} 
\caption{
{\bf Effect of heterogeneous tree activation.} (A) Response functions
$F(h_0)$ for the exponential activation distribution $h(g)=h_0
\exp(ag)$ where $g$ refers to the branchlet generation and $a$
controls the exponential shape. More distal branchlets (larger $g$)
are more activated than the apical site. For large values of parameter
$a$ the sensitivity of the response function is greatly increased
while the saturation remains almost the same. All curves have
$p_\lambda=0.8$, $p_\gamma=0.5$ and $\beta=1.0$. (B) Dynamic range for
the previous case with $h(g)$ activation, with an amazing enlargement
of the dynamic range for $p_\lambda=0.5$ and $\beta=1.0$. All cases
refer to tree sizes of $G=10$. (C) Response functions $F(h_0)$ for the
disordered branchlet activation model with coefficient of variation
$\kappa = 0.5$, recovery probability $p_\gamma = 0.5$, symmetric
propagation ($\beta = 1$) and $G = 10$. From bottom to top,
$p_\lambda=0, 0.2, \ldots,1$. (D) The dynamic range remains the same
for this disordered scenario in the tree (same parameters of panel
(C)). Note that a coefficient of variation $\kappa = \sigma / h_0 = 1$
corresponds already to a highly heterogeneous case.  }
\label{fig:quenched} 
\end{figure}

%\section*{Tables}
%\begin{table}[!ht]
%\caption{
%\bf{Table title}}
%\begin{tabular}{|c|c|c|}
%table information
%\end{tabular}
%\begin{flushleft}Table caption
%\end{flushleft}
%\label{tab:label}
% \end{table}

\end{document}